%% file: main.tex
\documentclass[twocolumn]{aastex631}

\usepackage{lineno}
\usepackage{CJK}
\begin{document}
\begin{CJK*}{UTF8}{gbsn}
\title{Simulating continuum-based redshift measurement in the \textit{Roman's} High Latitude Spectroscopy Survey}

\input{author.tex}



\begin{abstract}

We investigate the capability of  the \textit{Nancy Grace Roman Space Telescope's (Roman)} Wide-Field Instrument (WFI) G150 slitless grism to detect red, quiescent galaxies based on the current reference survey.
%We investigate the redshift recovery efficiencies for quiescent galaxies with the \textit{Nancy Grace Roman Space Telescope's (Roman)} Wide-Field Instrument (WFI) G150 slitless grism. 
We simulate dispersed images for \textit{Roman} reference High-Latitude Spectroscopic Survey (HLSS) and analyze two-dimensional spectroscopic data using the grism Redshift and Line Analysis (\verb|Grizli|) software. This study focus on assessing \textit{Roman} grism's capability for continuum-level redshift measurement for a redshift range of $0.5 \leq z \leq 2.5$. The redshift recovery is assessed by setting three requirements of: $\sigma_z = \frac{\left|z-z_{\mathrm{true}}\right|}{1+z}\leq0.01$, signal-to-noise ratio (S/N) $\geq 5$ and the presence of a single dominant peak in redshift likelihood function. We find that, for quiescent galxaies, the reference HLSS can reach a redshift recovery completeness of $\geq50\%$ for F158 magnitude brighter than 20.2 mag. We also explore how different survey parameters, such as exposure time and the number of exposures, influence the accuracy and completeness of redshift recovery, providing insights that could optimize future survey strategies and enhance the scientific yield of the \textit{Roman} in cosmological research.

%We investigate the redshift recovery capabilities for red galaxies within the context of the Roman Space Telescope's High Latitude Spectroscopy Survey using grism simulations. The simulations leverage the Grism Redshift and Line Analysis (\verb|Grizli|) software to generate and analyze two-dimensional spectroscopic data, focusing on the robustness of redshift estimations across varying exposure times and observational conditions. This study assesses the potential of the Roman grism for continuum-level redshift measurement. We systematically explore how different simulation parameters, such as exposure time and the number of exposures, influence the accuracy and completeness of redshift recovery. The results underscore the critical balance between exposure time and redshift recovery efficiency, providing insights that could optimize future survey strategies and enhance the scientific yield of the Roman Space Telescope in cosmological research.

\end{abstract}
\keywords{Galaxy spectroscopy; Astronomical simulations; Quiescent galaxies; High-redshift galaxies}
%%%%%%%%%%%%%%%%%%%%%%%%%%%%%%%%%%%%%%%%%%%%%%%%%%%%%%%%%%%%%%%%%%%%%%%%

\section{Introduction} \label{sec:intro}
%%%%%%%%%%%%%%%%%%%%%%%%%%%%%%%%%%%%
% -Importance of red galaxies for cosmology
% -Roman Grism detecting red galaxy in NIR
%%%%%%%%%%%%%%%%%%%%%%%%%%%%%%%%%%%%

Red, quiescent galaxies represent a fundamental component of the cosmic landscape, playing a pivotal role in the study of cosmological evolution and structure formation. These galaxies, characterized by their lack of significant star formation and their red colors due to older stellar populations, serve as important tracers of the mass assembly history of the universe~\citep{Slob24,Choi14,Beverage24,Zhuang23,Khullar22,Marsan22}. Understanding the distribution and properties of red, quiescent galaxies across cosmic time provides insights into the mechanisms that drive galaxy evolution and the influence of environment on these processes. In cosmology, red, quiescent galaxies are often utilized as proxies for identifying high-density environments like galaxy clusters and groups. The presence of the 4000\AA~break in their spectra makes them ideal candidates for accurate redshift estimation. Due to their brightness and distinct spectral features, these galaxies are easily detectable over a wide range of redshifts, making them ideal probes for studying the large-scale structure of the universe, such as with Baryon Acoustic Oscillation (BAO)~\citep{DESI3,Percival07,Rosell22,Eifler21} and Galaxy Clustering (GC)~\citep{Padmanabhan07,Pandey022,Yuan24,Sailer24,White22}.

Based on the current reference survey design, the \textit{Nancy Grace Roman Space Telescope (Roman)} will conduct an infrared (IR) survey over a $\sim$2000 deg$^2$ area as part of the High Latitude Wide Area Survey \citep[HLWAS;][]{Spergel15, Wang22} to enable weak lensing and clustering measurements. During its 5-year prime mission, \textit{Roman} will conduct the HLWAS which includes an imaging component (High Latitude Image Survey; HLIS) and a spectroscopic component (High Latitude Spectroscopic Survey; HLSS). 

The HLSS has a slitless grism component. In particular, the \textit{Roman} grism enables spectroscopy with resolution of 600 over wavelength range of $\lambda = 1 - 1.93\mu m$, which is capable of probing red, quiescent galaxies at high redshift by capturing the 4000 Å break redshifted into the coverage of the grism. While the primary target of the \textit{Roman} grism is emission line galaxies (ELGs), specifically those with H$\alpha$ and [OIII] lines, which are critical for BAO and Redshift Space Distortion (RSD) measurements~\citep{Eifler21}, it is also worth noting that several previous studies \citep[see, e.g.,][]{Ryan07, Joshi19, Eugenio21} have shown that continuum-based redshifts derived from low-resolution grism spectroscopy and photometry based on 4000\AA\ or Balmer breaks can also provide reasonably accurate redshift determinations (on the order of ${\sim}$1-2\%). One of the earlier works by \citet{GladdersYee2000} showed that a ${\sim}10$\% redshift accuracy can be expected for galaxies in clusters in certain redshift ranges with just a color cut using two filters. % This method of course relies on the 4000\AA~break being sampled by the observing filters. A low-resolution grism spectrum simply extends this idea to ....
\citet{Joshi19} also showed that the expected number of red galaxies exhibiting a 4000\AA~break is comparable to ELGs with prominent emission lines, albeit with higher redshift uncertainty, allowing for the analysis of a much larger sample relative to a sample with just ELGs. Furthermore, previous studies have confirmed the presence of red, quiescent galaxies at high redshifts up to $z\ {\approx}\ 4$ with VANDELS survey~\citep{Carnall19, Carnall20}, the Hubble Space Telescope grism WFC3/G141~\citep{Whitaker13,Eugenio21} and the James Webb Space Telescope (JWST) NIRSpec spectroscopy \citep{Nanayakkara24,Carnall24}. Consequently, the \textit{Roman} grism could potentially yield a substantial amount of information about red, quiescent galaxies at $z > 1$ given its NIR coverage and wide survey area of $2000 \deg^2$. Depending on the sample size, this dataset could be used independently or combined with other samples to enhance analyses, such as those for BAO and galaxy clustering.

Efforts have been make to assess the performance of the spectroscopic components of \textit{Roman}. For example,~\citet{Wang22} simulated \textit{Roman} grism images based on a proposed reference survey design, ~\citet{Bhavin22} simulated \textit{Roman} P127 prism observations to quantify the efficiency of recovered Type Ia supernova redshifts as a function of exposure time, and ~\citet{Wold23} simulated and proposed an observing strategy for the \textit{Roman} grism deep fields to study high redshift Lyman-$\alpha$ emitters.
This paper primarily focuses on assessing the impact of exposure time on the redshift recovery efficiency of red galaxies with the \textit{Roman} grism. We use simulated direct images for \textit{Roman} HLIS from~\citet{Troxel22} as input to simulate and analyze two-dimensional (2D) spectroscopy images through the \textit{Roman} grism with the Grism Redshift and Line Analysis software, \verb|Grizli|\footnote{\url{https://github.com/gbrammer/grizli}}~\citep{Grizli}. Our study specifically targets the redshift estimation of red, quiescent galaxies in these simulations. We estimate redshifts using two separate packages -- \texttt{Grizli} and \texttt{Bagpipes}\footnote{\url{https://bagpipes.readthedocs.io/en/latest/}} \citep{Carnall18, Carnall19}. This paper is structured as follows: Section~\ref{sec:Grism-sim} describes the simulation tool and inputs used to simulate \textit{Roman} grism images. In Section~\ref{sec:method}, we provide an overview of \verb|Grizli|'s contamination modeling and redshift fitting pipeline. Section~\ref{sec: gal selection} details our selection criteria for choosing red galaxies with accurate redshift estimations in our simulations. We present and discuss the results in Section~\ref{sec:results}. Finally, we summarize our findings.

%%%%%%%%%%%%%%%%%%%%%%%%%%%%%%%%%%%%%%%%%%%%%%%%%%%%%%%%%%%%%%%%%%%%%%%%
\section{Grism Simulation Inputs} \label{sec:Grism-sim}
%%%%%%%%%%%%%%%%%%%%%%%%%%%%%%%%%%%%%%%
% -Grizli overview
% -Direct imaging data
% -Input SEDs
% -Roman Grism overview + survey stradegy + simulation specs (e.g BG level, etc..)
% -Caveats (mention limitations)
%%%%%%%%%%%%%%%%%%%%%%%%%%%%%%%%%%%%

\begin{figure}[t]
    \centering
    \includegraphics[scale = 0.5]{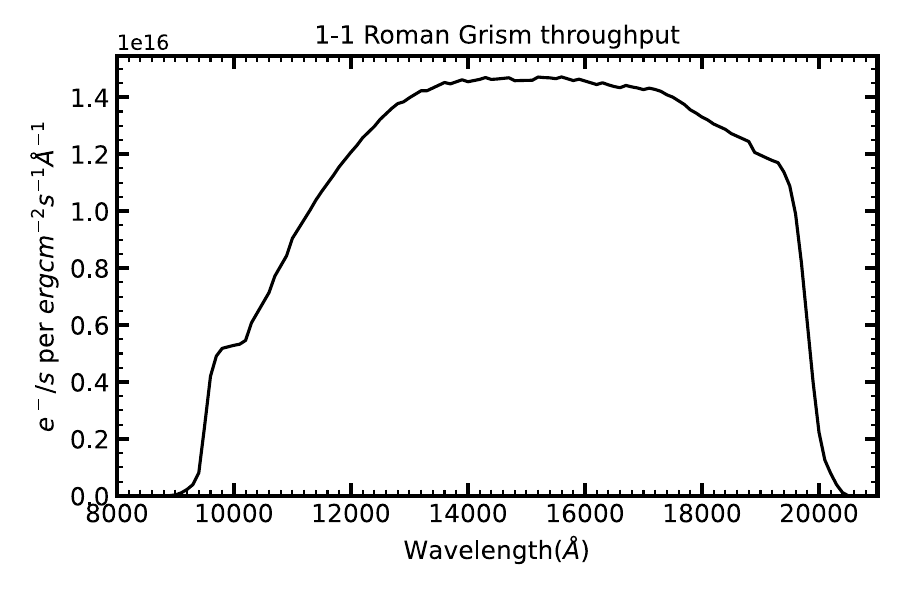}
    \caption{Roman WFI grism sensitivity functions for the first spectral order (1-1).}
    \label{fig:grism-throughput}
\end{figure}

For simulating realistic grism images for the \textit{Roman} HLSS, we utilize the images produced by \citet{Troxel22}. These images are then processed using the Grism Redshift and Line Analysis software, \verb|Grizli| \citep{Grizli}, which provides a comprehensive end-to-end simulation and processing pipeline for analyzing grism data. Simulating the grism image of a given field requires following inputs:  1) Direct images and corresponding segmentation maps, along with catalogs with information for all the objects including their segmentation id, ra, dec, magnitude, redshift, 2) spectral energy distributions (SEDs) for all objects, 3) the configuration file that describes the spectral trace and dispersion for the \textit{Roman} grism along with a corresponding sensitivity curve. In this paper we use the default \textit{Roman} grism configuration generated by \verb|Grizli|. For this study, we exclusively focus on the first spectra order (science order) and use the grism sensitivity curve as employed by by~\citet{Wang22} as in shown in Fig~\ref{fig:grism-throughput}. We will discuss 1) and 2) in more details in the following sections. 

%%%%%%%%%%%%%%%Figure%%%%%%%%%%%%%%%%%
%\begin{figure*}[!t]
%    \centering
%    \includegraphics[scale = 0.6]{sim_exmaple.pdf}
%    \caption{An example of the input direct noiseless image and simulated grism image with %noise added.}
%    \label{fig:direct-grism-plot}
%\end{figure*}
%%%%%%%%%%%%%%Figure%%%%%%%%%%%%%%%%%%

\subsection{Direct images}

\begin{figure*}[!t]
    \centering
    \includegraphics[scale = 0.6]{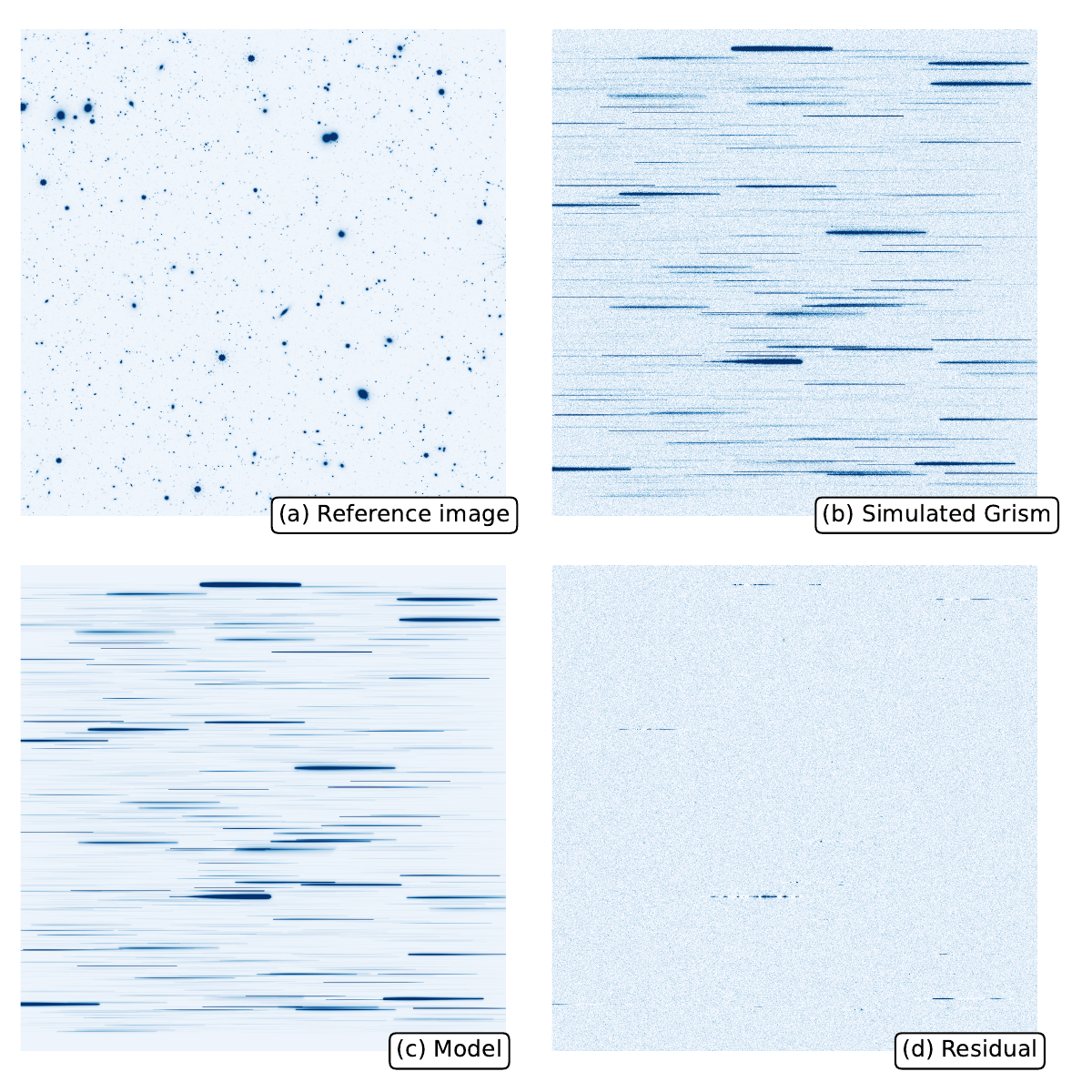}
    \caption{Spectral contamination model of Roman grism simulation. The panels show: (a) the reference image, the same direct image used to simulate the grism image. (b) the simulated grism image (c) the spectral contamination model derived from the direct image and polynomial fits for the spectra for all objects. (d) the residuals after subtracting the contamination model from the simulated grism image.}
    \label{fig:global-contam-example}
\end{figure*}

We use simulated noiseless F158 direct images from~\citet{Troxel22}. This simulation constitutes a 20 deg$^2$ of overlapping synthetic imaging surveys representing the full depth of the Nancy Grace \textit{Roman} HLIS along with five years of observations of the \textit{Vera C. Rubin Observatory Legacy Survey of Space and Time (LSST)}. We refer the reader to~\citet{Troxel22} for more details. We use noiseless images to prevent the propagation of noise from the direct images through the dispersed direction in the simulated grism images, which would result in inaccurate electron count number for the spectra. Additionally, we have experimented with using coadded images instead of noiseless images and found that the results are comparable. The corresponding segmentation maps are generated using Source Extractor~\citep{SExtractor}. The object catalog, which encompasses positional and segmentation information for all detected objects, was generated by cross-referencing the output from Source Extractor with the input truth catalog provided by~\citet{Troxel22}. Specifically, the  process involves matching objects based on their celestial coordinates, which is then further refined by selecting those with the closest photometry. The Roman Wide-Field Instrument has 18 4096×4096 Sensor Chip Assemblies (SCAs)\footnote{\label{footnote:roman-tech}\url{https://roman.gsfc.nasa.gov/science/WFI_technical.html}}. In this paper, we selected F158 band noiseless single exposure images from two random telescope pointing. For each telescope pointing, we generated simulated grism images for all 18 sensors. In this study We work with image in unit of e$^-$/s.

\subsection{Input Spectral Energy Distributions (SEDs) }
The expected $5\sigma$ limit of \textit{Roman} grism for continuum sources over a one-hour exposure is approximately 20.5 AB magnitude, although this value may vary depending on the wavelength\footref{footnote:roman-tech}. We simulate all galaxies in the direct image down to $m_{F158} = 26$, which is more than enough to cover the expected depth of detecting red galaxies, in order to account for the background contamination. 

To have statistically significant sample size for quantifying redshift efficiency for red galaxies, we assign old, quiescent galaxy spectra to a fraction of galaxies within the redshift range of 0.5 to 2.5 and with $m_{F158}$ magnitudes less than 22.5 in each image. The rationale for selecting the magnitude cut-off is explained in Section~\ref{sec: gal selection}, where we demonstrate that the current reference HLSS is capable of achieving reliable redshift measurements for red, quiescent galaxies only up to a magnitude of 21. 

This quiescent spectra assignment was made at a rate of 60\%, which is designed to guarantee a statistically significant red galaxy sample size while ensures a representation of a diverse range of other galaxy types in both foreground and background, thereby preventing an oversimplification of the galaxy populations in each simulated image. These quiescent galaxy spectra are generated by using \verb|python-fsps| package \citep{Conroy09, ConroyGunn10, Johnson2024-fsps}. In terms of other galaxies, the original input SEDs \citet{Troxel22} used are generated based on a limited number of SED templates and do not contain emission line features~\citep{Cosmodc2}. Incorporating emission line sources into the simulation is crucial, as they are primary contributors to contamination during spectra extraction and redshift estimation processes. Including these sources ensures that our simulation and analysis closely resembles those of the actual observations. Thus, for those galaxies, we use the best-fit EAZY SEDs from the 3D-HST survey~\citep{Momcheva16} and randomly assigned SEDs to these galaxies. For stars, we use spectra from the Pickles stellar library~\citep{Pickles98}. All input SEDs are normalized to the direct imaging F158 bandpass before they are fed into the simulation.

\subsection{Position angles and Background} \label{subsec:sim config}
For each selected pointing, we generate \textit{Roman} grism images across all 18 WFI detectors at four roll angles (position angles, PA) —0, 5, 170, and 175 degrees following the reference HLSS design~\citep{Wang22}. For each roll angle, we simulate two exposures, each with an exposure time of 347 seconds\footnote{The exposure time is based on \url{https://roman.ipac.caltech.edu/sims/IPAC-STScI_Goddard_Grism_sim.html}}. However, we note that we do not simulate the exact dithering and tilting pattern since our simulations are simply designed to evaluate the impact of exposure time on redshift recovery rate of red galaxies. Instead, all the roll angles share the same field center. Thus, our simulations assumes that all sources appear in same the place on the detector and are observed in all 8 grism observations, resulting in a total exposure time of 2776 seconds for each source. According to~\citet{Wang22}, the current observing plan yields a coverage of 487, 1162, 1712, 1961, 2046, 2076, 2102, and 2117 deg$^2$ with coverages of $>=$ 8, 7, 6, 5, 4, 3, 2, and 1 grism observation, respectively. We investigate the impact of varying these exposure times on quiscent galaxies redshift recovery rate in Section~\ref{sec:impact of PA}. Our grism simulation contains three noise components, a zodiacal background of 1.047 e$^-$/pix/s and a dark current of 0.0015 e$^-$/pix/s, following the noise model of~\citet{Bhavin22}, with a readout noise rms of 16 electrons\footnote{\url{https://roman.gsfc.nasa.gov/science/technical_resources.html}}. The simulated image is stored as \verb|.flt| file\footnote{\url{https://grizli.readthedocs.io/en/latest/api/grizli.model.GrismFLT.html}} for each PA. Panel (a) and (b) in Figure~\ref{fig:global-contam-example} show an example of an input noiseless direct image and corresponding simulated \textit{Roman} grism images. 

\subsection{Caveats}
As mentioned in the earlier section, we focus exclusively on simulating first-order (1-1) spectra. It's important to recognize that the \textit{Roman} grism instrument disperses light into multiple spectral orders, with the first order (1-1) being the primary focus for scientific analysis, while the others are considered unwanted orders. Our decision to overlook spectra from orders other than the science order may influence the outcomes of spectral extraction and redshift determination for two primary reasons. First, this approach implicitly assumes that all light captured by the \textit{Roman} Grism is attributed to the first order, which simplifies the actual process and may lead to overestimated Signal-to-Noise Ratio (S/N) of the extracted spectra. Second, the zeroth order will appear as a compact spot on the grism image, which can be confused with emission lines, particularly in the case of bright objects. Such misidentification, which is not accounted for in our simulation, could potential further contaminate the extracted spectra besides overlapping spectra, leading to inaccuracies in the subsequent redshift fitting. However, a recent study on the spectral characterization of the \textit{Roman} grism by \citet{Bray24} shows that the power intensity of the unwanted orders relative to the science (1-1) order is only a few percent or even less. Therefore, we anticipate that the most of the contamination from unwanted orders in the \textit{Roman} grism will be minimal, provided there are no extremely bright sources near the target galaxy.

Additionally, the input direct image is pre-configured with an embedded PSF as we use the F158 band direct images as the simulation input. Nevertheless, due to the  differences in the profiles of the F158 and \textit{Roman} Grism bandpass, the embedded PSF may not serve as an accurate approximation. Furthermore, our simulations do not account for the variation in the point spread function (PSF) with wavelength. Lastly, the noise model employed in our simulations is basic, incorporating only zodiacal background, dark current, and readout noise, assuming other potential sensor anomalies, such as the impact of cosmic rays, have already been removed. All simulated images are assumed to be flat-fielded. While we believe such simplifications are sufficient for the goal of this study, future simulations are needed to accurately quantify the impacts of these effects.
%%%%%%%%%%%%%%%%%%%%%%%%%%%%%%%%%%%%
\section{Analysis method} \label{sec:method}
%%%%%%%%%%%%%%%%%%%%%%%%%%%%%%%%%%%%%%%
% Spectra extraction
% Contanmination modelling 
% SED fitting + Redshift fitting
% Redshift fitting cross-check with other alogrithm
%%%%%%%%%%%%%%%%%%%%%%%%%%%%%%%%%%%%%%%
\begin{figure}[!t]
    \centering
    \includegraphics[scale = 0.7]{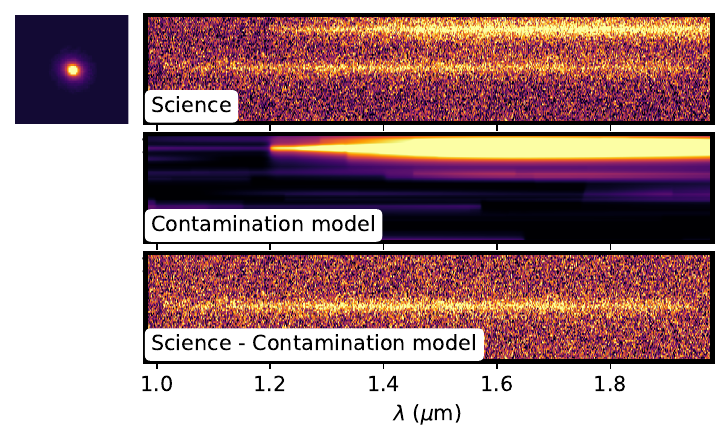}
    \caption{A single 2D cutout region of the source in one position angle (PA = 0 degrees). The target galaxy is shown in the upper left. Top panel: cutout region of the simulated Roman grism images centered on the source.  Middle panel: Contamination model obtained by iteratively fitting polynomial spectral templates to the grism spectra of all objects in the field. Bottom panel: Cleaned science image, on which the redshift fitting is performed, after subtracting the contamination model.}
    \label{fig:extrac_contam_ex}
\end{figure}

Upon completing the simulations, we proceed to utilize \verb|Grizli| for the extraction of spectra and the estimation of redshifts. \verb|Grizli| is employed to derive the two-dimensional grism data from selected sources and to conduct redshift fitting. It has the functionality to combine multiple exposures from different grisms and/or different roll angles to analyze the spectrum of a given object. In this study, we use the default contamination modeling and redshift fitting scheme implemented by \verb|Grizli|. Here, we give a brief overview of its spectrum analysis pipeline and refer the readers to the official documentation for further details. 

\begin{figure*}[!h]
    \centering
    \includegraphics[scale = 0.6]{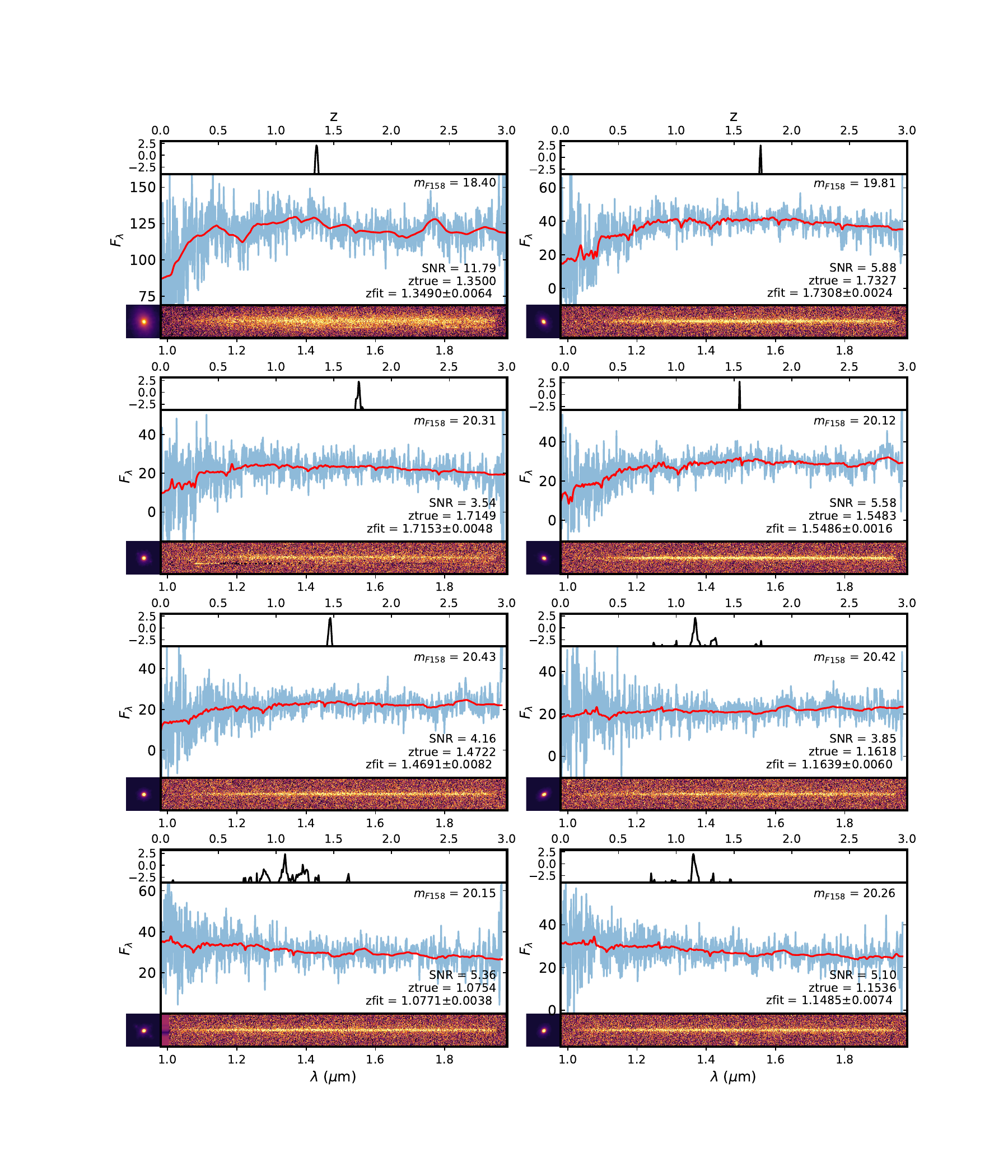}
    \caption{Example spectra for eight red galaxies randomly selected from the simulation. \textit{Upper panels:} redshift probability distribution, $log[P(z)]$, obtained from initial coarse redshift fitting phase for each target. \textit{Middle panels:} Optimally extracted 1-D spectra (blue) derived by combining all exposures of each target and corresponding Grizli best-fit templates (red). The flux unit is in $10^{-19}$ erg/s/cm$^2$/$\AA$. The S/N, true redshift and best-fit redshift are presented in the lower right legend. The F158 magnitude of each galaxy in direct images is shown in upper right legend. \textit{Lower panels:} corresponding  direct images and simulated 2D G150 spectra (PA = $0^{\circ}$).}
    \label{fig:extraction_example}
\end{figure*}

\begin{figure*}[t]
    \centering
    \includegraphics[scale=0.6]{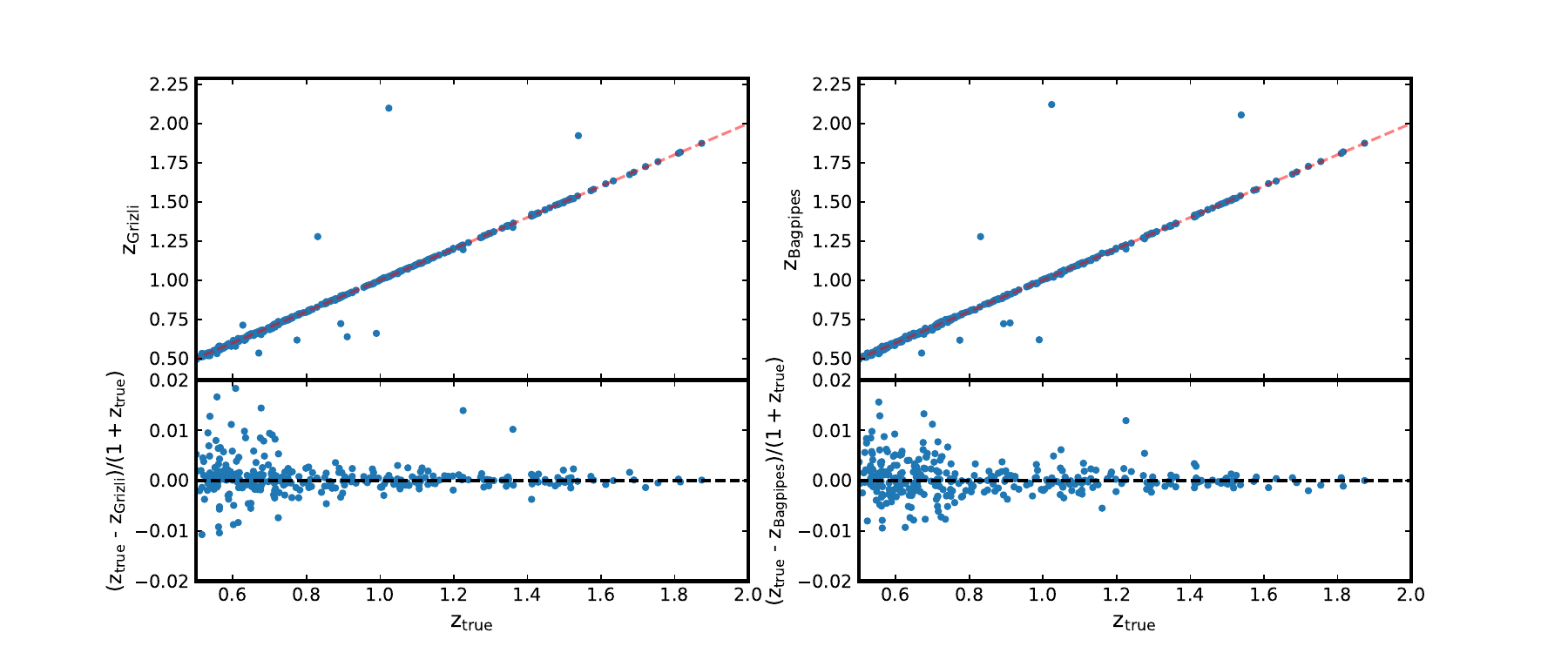}
    \caption{Comparison of true redshifts versus measured redshifts using Grizli (left panel) and Bagpipes (right panel) for all galaxies with a signal-to-noise ratio (S/N) greater than 5 in the simulation. The bottom panels show the residuals, defined as $(z_\mathrm{fit} - z_\mathrm{true})/(1 + z_\mathrm{true})$ as a function of true redshift. Note that the plotting range for the residual plots has been adjusted to better highlight the residuals, resulting in the exclusion of outliers from the upper panels.}
    \label{fig:z_grizli_vs_bagpipes}
\end{figure*}

\subsection{Contamination model}

Since slitless spectroscopy captures the spectra of all sources in the images simultaneously, the target spectrum can be heavily contaminated by overlapping spectra from neighboring sources. Therefore, it is essential to have a model that accurately characterizes any potential contamination. This spectral contamination model can then be subtracted from the science image, resulting in a clearer view of the target spectrum.

The grism images are stored as \verb|.flt| file for each PA simulated. First, \verb|Grizli| initializes a \verb|GroupFLT| class to provide a container for processing multiple \verb|.flt| exposures simultaneously. The next step is to model spectra continua of all sources in the field to build a global spectral contamination model for each \verb|.flt| image. This is done in two passes. A first-pass model is computed for objects assuming a simple flat continuum. This step is used to account for the pixel space taken by each source and therefore helps identify the pixels of a given object that could be contaminated by neighbouring sources. In the second pass, a refined model based on a third-order polynomial fit is fitted to every spectrum after the initial contamination model from neighboring sources is subtracted. This refined model is then used to update the original contamination model. The second step is iterated three times by stepping through all objects starting from the brightest object in the field. Panel (b) - (d) in Figure~\ref{fig:global-contam-example} show an example to illustrate this process for a field picked in our simulation. %The spectral continua of the sources are modelled using an iterative polynomial fitting of the data for contamination estimate and removal.

\subsection{Redshift fitting \& Spectral extraction} \label{subsec:redshift fitting}

The determination of redshift is carried out through a two-step fitting process. Initially, a preliminary fit is applied using a coarse redshift grid ($dz/(1+z) \approx 0.005$), utilizing a predefined set of continuum templates and line complex templates available in \verb|Grizli|. This is followed by a refined fitting procedure, focused around the chi-squared minimum derived from the initial coarse fit with a finer redshift grid to resolve the best redshift fit. 
\verb|Grizli| extracts the contamination-subtrated 2D cutout grism frame for each sources in the field, as illustrated in Figure~\ref{fig:extrac_contam_ex}, and the redshift fitting is performed by modelling the templates on to those frames. The advantage of doing this is that it helps account for the spectral smearing and broadening of pixles in direct images, also known as self-contamination, of the grism data. We perform fitting to all red galaxies in our simulation in a redshift range of $\mathrm{z} = 0 - 10$.

Even though the redshift fitting is performed on 2D cutout images, \verb|Grizli| still provides the functionality to extract the 1D spectrum by combining multiple exposures from all position angles based on the optimal extraction algorithm from~\citet{Horne86}. We use these extracted 1D spectra to visualize the fitting results and to calculate the signal-to-noise ratio (S/N) of each source. The S/N are computed over the entire optimally extracted 1-D spectra using the DER-SNR algorithm by~\citet{Stoehr08}. We present several examples in Figure~\ref{fig:extraction_example}, which shows the redshift likelihood function P(z) from the initial coarse redshift fitting phase in the top panels, the optimally extracted 1-D spectra alongside their corresponding best-fit templates in the middle panels, and 2D cutouts of the simulated grism image with corresponding direct imaging in the bottom panel. We note that even if the optimal weighting technique introduced by~\citet{Horne86} is not directly applicable to slitless spectroscopy~\citep{Kummel09},  we still achieve reasonable redshift fitting results from optimally extracted 1D spectra (see Section~\ref{subsec: Grizli_vs_Bagpipes}). Additionally, the S/N derived from the extracted 1D spectra nonetheless proves to be a reliable indicator of identifying quiescent galaxy candidates with accurate redshift estimations (see Section~\ref{subsec:SNR cut}).

\begin{figure}[h]
    \centering
    \includegraphics[scale = 0.58]{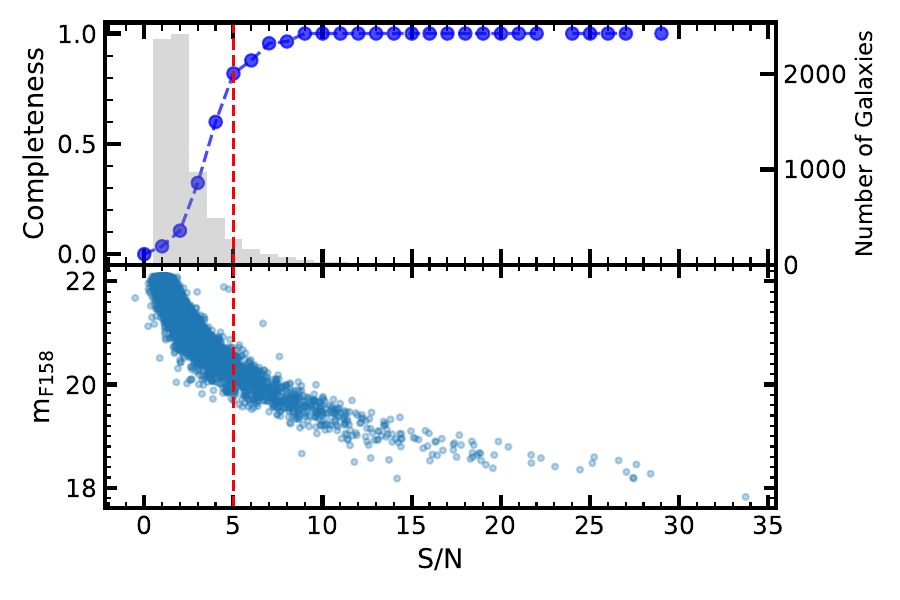}
    \caption{Top panel: Redshift completeness,  defined as the fraction of galaxies with reliable redshift estimates (single dominant peak in P(z)) and small residuals ($\sigma_z \leq 0.02$), for red galaxies versus S/N. The red dashed line indicates the S/N threshold of 5, beyond which $90\%$ completeness is achieved. Bottom panel: F158 band magnitude as a function of S/N for the entire galaxy sample in the simulation.}
    \label{fig:SNR_verify}
\end{figure}

\begin{figure*}[!t]
    \centering
    \includegraphics[scale = 0.65]{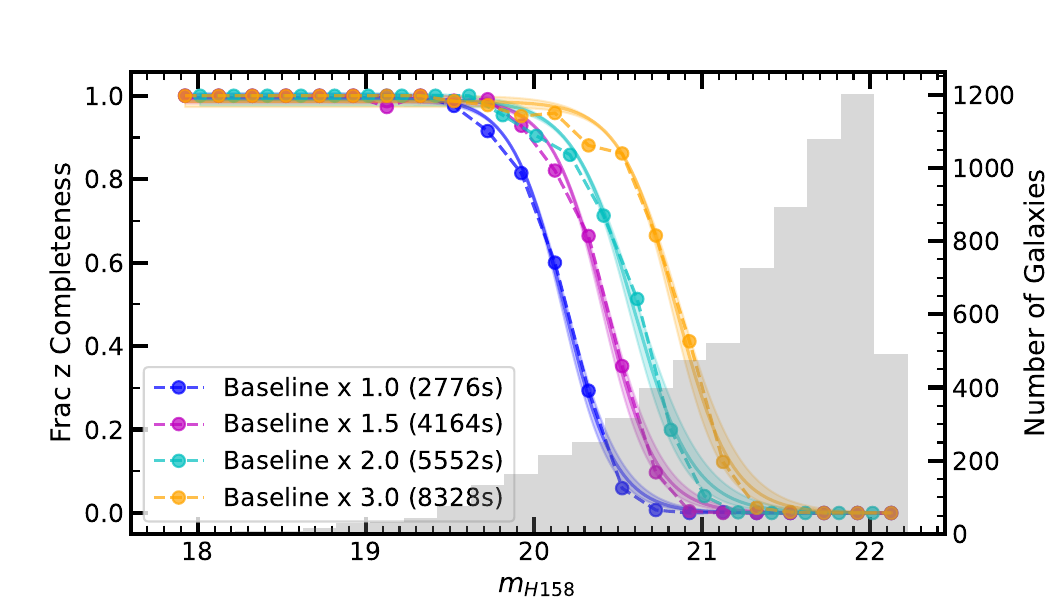}
    \caption{Efficiency of Redshift Recovery by Galaxy Magnitude in F158 across 36 Simulated Grism Images with different exposure times. These simulations assume that objects are observed in all 8 exposures. The left axis indicates redshift completeness, representing the percentage of galaxies in each magnitude bin meeting specific criteria: 1) Redshift accuracy with $\sigma_z = \left|\Delta z/(1+z)\right| \leq \textbf{0.01}$, 2) SNR $\geq \textbf{5}$, and 3) Presence of a singular peak in the redshift PDF. The histogram in the background illustrates the distribution of galaxy magnitudes among the simulated sample, with the number of galaxies displayed on the right axis. Data points reflect the observed redshift completeness, with various colors denoting different exposure times as detailed in the legend. Solid lines represent logistic regression models fitted to the observed redshift completeness data.}
    \label{fig:zeff_exptime}
\end{figure*}

\subsection{Verficaition of Grizli redshift estimates} \label{subsec: Grizli_vs_Bagpipes}
As discussed in the previous section, redshift estimation using \verb|Grizli| is based on 2D spectra. However, to ensure broader applicability and to engage the interest of the community, it is crucial to confirm that similar results can be achieved using 1D spectra. Additionally, corroborating these results with an independent redshift estimation package reinforces the reliability of the fitting algorithm implemented by \verb|Grizli|. To validate Grizli's redshift estimations, we compare its results with those obtained from the Bayesian Analysis of Galaxies for Physical Inference and Parameter Estimation (\verb|Bagpipes|) package\footnote{\url{https://bagpipes.readthedocs.io/en/latest/}}, which performs spectroscopic redshift fitting based on 1D spectra. For this validation, we select all quiescent galaxies from our simulation with S/N greater than 5, and use optimally-extracted 1D spectra as input for Bagpipes. For a detailed description and documentation of Bagpipes, we refer readers to \citet{Carnall19,Carnall20}. The results are shown in Figure~\ref{fig:z_grizli_vs_bagpipes}, where the true versus measured redshifts using \verb|Grizli| and \verb|Bagpipes| are shown in the left and right panels, respectively. The comparison demonstrates a strong agreement between the two independent packages.

\section{Candidate selection} \label{sec: gal selection}
In the actual survey, where truth information is not accessible to determine the reliability of the spectra extraction and to evaluate the accuracy of redshift measurement, we need to rely on other available information to help decide if a galaxy has accurate redshift measurement and if it is a good quiescent galaxy candidate. We therefore apply the following thresholds to selection good candidates from our simulation.  

\subsection{Redshift likelihood}
As outlined in Section~\ref{subsec:redshift fitting}, redshift fitting is conducted in two stages. To eliminate ambiguous redshift determinations, a galaxy is only considered a suitable candidate if there is a single dominant peak in the redshift likelihood distribution, P(z), during the initial coarse fitting phase. This helps ensure that the redshift estimates are accurate and reliable, minimizing the risk of selecting galaxies with incorrect or uncertain redshift measurements. We find that filtering galaxies in this way always leads to a sample with the most accurate redshift estimations.

\subsection{Signal-to-noise ratio} \label{subsec:SNR cut}
We also examine the relation between S/N and the goodness of redshift estimation. To achieve this, we simulate grism images based on the strategy outlined in Section~\ref{subsec:sim config}, assuming the reference survey~\citep{Wang22} exposure time, and analyze the extraction and redshift fitting results. In this work, we use redshift residual as defined by $\sigma_z = (z_\mathrm{fit} - z_{\mathrm{true}})/(1 + z_\mathrm{true})$ to evaluate the goodness of redshift estimation. Figure~\ref{fig:SNR_verify} shows the fraction of red galaxies with a single dominant peak in P(z) and $\sigma_z \leq 0.01$, relative to total red galaxy population in each S/N bin, as a function of S/N. It achieves 90\% completeness at an S/N of approximately 5. We investigate galaxies exhibiting poor redshift estimates despite having high S/N ($\geq5$). Upon visual inspection of these outlier spectra, we identify that those inaccurate redshift measurements are due to either incomplete spectra, as some sources are near the sensor edge, or heavy contamination from a neighboring bright star, which our contamination model could not completely mitigate. Thus, we consider an S/N $\geq 5$ generally as an indicator for good spectral quality, provided that the spectra are not located near the sensor edge or affected by contamination from nearby bright stars.
 
We note that only $10\%$ galaxies in our simulated red galaxy sample has $\mathrm{S/N} \geq 5$. The F158 band magnitudes versus S/N for all the galaxies included in our simulation are shown in the bottom panel of Figure~\ref{fig:SNR_verify}. The S/N $\geq 5$ cut constrains the magnitude of our galaxy sample to $\mathrm{m}_{\mathrm{F158}} \approx 21$. This indicates that, although we simulate red galaxies with $m_{\mathrm{F158}} \leq 21$, reliable redshift estimates are not achievable for these targets given the exposure times of the reference survey. This justifies our choice of limiting the magnitude of simulated red galaxies to $m_{\mathrm{F158}} = 22.5$, which allows us to adequately cover the lower end of the redshift efficiency curve.

%%%%%%%%%%%%%%%%%%%%%%%%%%%%%%%%%%%%
\section{Results} \label{sec:results}

In this work, we select red galaxies  as suitable candidates for spectroscopy that satisfy the thresholds discussed above: a) a single dominant peak in the redshift likelihood function, b) S/N $\geq 5$, and c) $\left| \sigma_z \right| \leq 0.01$. The last threshold is applied specifically to rule out high bias cases, which often occur due to partially or heavily contaminated galaxy spectra, as outlined in Section~\ref{subsec:SNR cut}.
We define the red galaxy redshift completeness, or the redshift recovery rate, as the fraction of red galaxies meeting these criteria relative to the total simulated red galaxies population.

%\textbf{Outline:}
%\begin{enumerate}
%     \item Summary plot of 2D and 1D spectra
%     \item Impact of SNR on redshift fitting (justity the choice of $SNR > 3$)
%     \item Redshift recovery vs magnitude
%     \item Discussion on different Contanmination modelling stradgy
%\end{enumerate}

\subsection{Impact of exposure time on redshift efficiency}
To study the impact of exposure time on red galaxy redshift completeness, we repeatedly run \textit{Roman} grism simulation based on description of Section~\ref{subsec:sim config} for different exposure times. We treat the exposure time proposed for the HLSS reference survey as the baseline. In this section, we assume that all objects have grism spectra available in all 8 exposures. The impact of reduced number of available exposures is studied in the Section~\ref{sec:impact of PA}. Based on this assumption, we evaluate four distinct scenarios: the first employs a baseline exposure time, amounting to a total of $8\times347 = 2,776$ seconds. Subsequently, we extend this baseline by factors of 1.5, 2, and 3, resulting in total exposure times of 4,164 seconds, 5,552 seconds and 8,328 seconds respectively.

Figure~\ref{fig:zeff_exptime} shows the redshift completeness as a function of F158 magnitude in our simulation for the four exposure time schemes. Each magnitude bin shows the fraction of red galaxies, satisfying the three thresholds we set above in Section~\ref{sec: gal selection}, in the total red galaxy population in that bin. Figure~\ref{fig:zeff_exptime} shows that under baseline conditions, our simulation achieves a 50\% level for red galaxy redshift recovery rate at an approximate magnitude of 20.2. As exposure times per exposure are extended, the F158 magnitude at which 50\% completeness is attained increases to $m_\mathrm{F184} = $20.4, 20.6, and 20.9 for the three alternative scenarios, respectively.

\begin{figure}[t]
    \centering
    \includegraphics[scale = 0.5]{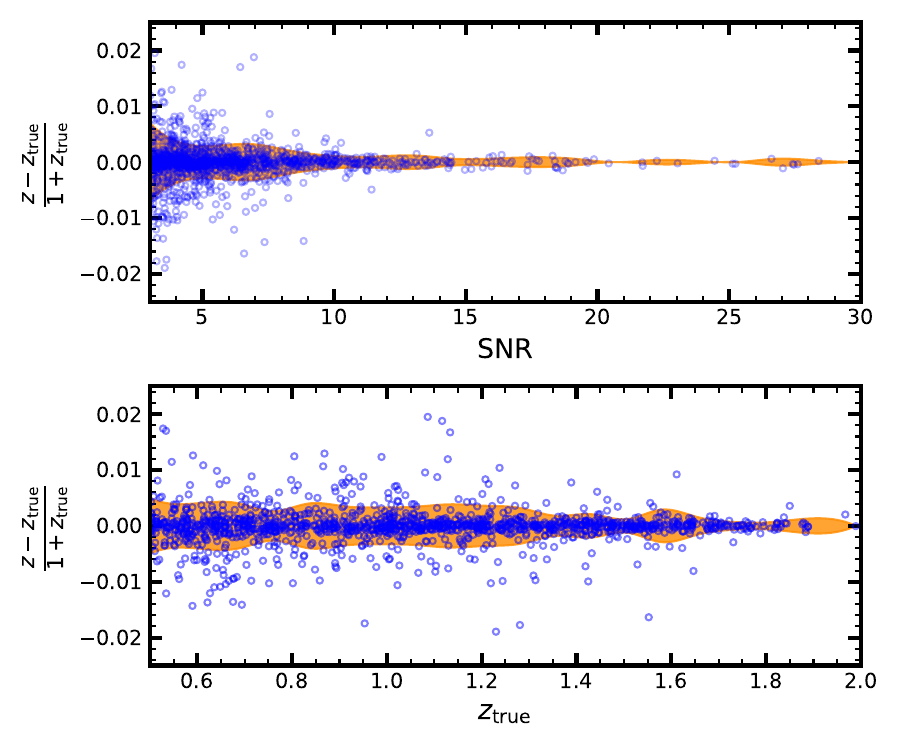}
    \caption{Residuals (z-$\mathrm{z}_{\mathrm{true}}$)/(1+$\mathrm{z}_{\mathrm{true}}$)  vs. SNR (upper panel) and true redshift (lower panel) based on standard exposure time in HLSS, assuming objects are covered in all 8 exposures. The orange shaded region shows the $\pm1\sigma$ region of the redisduals. All galaxies satisfying selection threshold are shown.}
    \label{fig:dz_vs_SNR_ztrue}
\end{figure}

Figure~\ref{fig:dz_vs_SNR_ztrue} shows the dependence of the redshift residual on S/N of the simulated spectrum and true redshift for all red galaxies that meet the specified thresholds in our simulation, specifically within the context of the baseline exposure time. It's clear that the accuracy increases with S/N. The bottom panel of Figure~\ref{fig:dz_vs_SNR_ztrue} suggests that higher redshift red, quiescent galaxies ($z_{\mathrm{true}}\geq1.4$) tends to have more accurate redshift estimation compared to those in lower redshift range. This can be attributed to the redshifting of the $4000\mathring{A}$ break, the most distinctive feature in the red galaxy spectrum, into the wavelength range covered by the \textit{Roman} grism, beginning at approximately $z = 1.4$. The presence of this spectral break help improves the accuracy of template fitting. This can be clearly seen in the top panels in Figure~\ref{fig:extraction_example}. Given comparable S/N, galaxies with $z_{\mathrm{true}} \geq 1.4$ clearly exhibit cleaner P(z) distributions during the initial coarse fitting phase compared to those at lower redshifts. It is also worth noting that reasonable continuum-based fits can still be achieved for sources with $z < 1.4$, even in the absence of the $4000\mathring{A}$ break.

\subsection{Impact of number of exposure on redshift efficiency} \label{sec:impact of PA}

\begin{figure}[t]
    \centering
    \includegraphics[scale = 0.5]{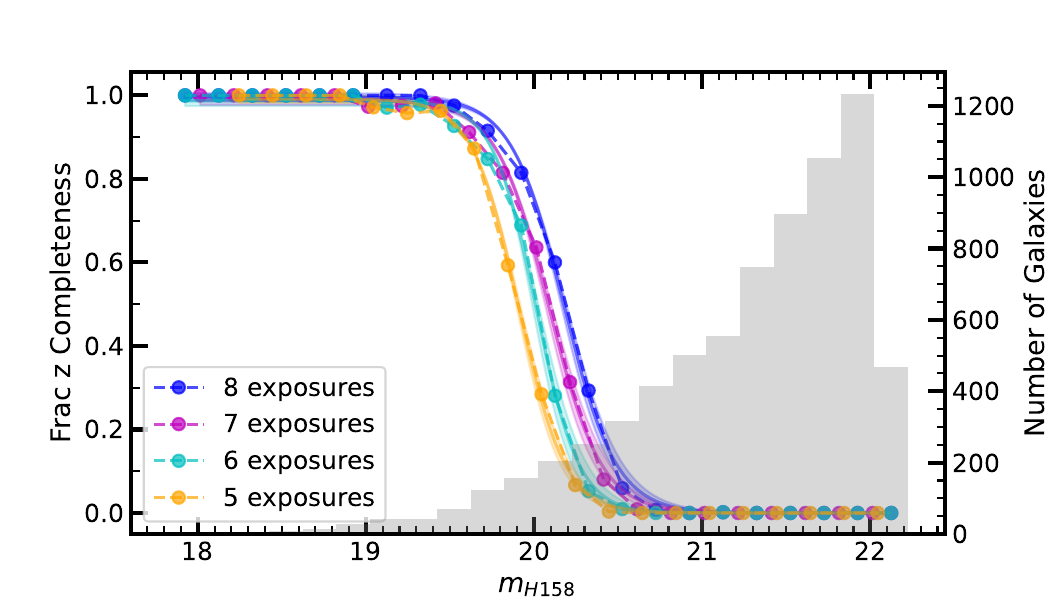}
    \caption{Same plot as Fig~\ref{fig:zeff_exptime} but for different number of available exposures.}
    \label{fig:zeff_PA}
\end{figure}

Based on the reference survey, only $23\%$ (487 out of 2117deg$^2$) of the total survey area will be observed in all eight exposures. Consequently, we also explore the effects of a limited number of grism exposures on the red galaxy redshift completeness. Specifically, we assess scenarios where for each objects only seven ($55\%$ of survey area), six ($80\%$ of survey area) and five ($93\%$ of survey area) grism exposures are available given baseline exposure time. This was achieved by randomly omitting one, two and three simulated exposures during the processes of 2-D spectrum cutout extraction and redshift fitting. Figure~\ref{fig:zeff_PA} illustrates the redshift completeness curves for these three scenarios, with the eight-exposure case included for comparison.The figure indicates that for each reduction in the number of exposures, there is an approximate 0.1 magnitude decrement in the F158 magnitude corresponding to a $50\%$ efficiency level.

\section{Summary \& Discussion} \label{subsec:summary}
We present the procedure to use \verb|Grizli| for simulating, extracting and fitting \textit{Roman} Grism spectra based on the HLSS reference survey design. We have reported results from the analysis of 7,227 simulated \textit{Roman} Grism red galaxy spectra with exposure times of 2770, 4164, 5552, 8328 seconds. We expect the redshift efficiency of red galaxies to be $50\%$ at approximately $m_{\mathrm{F158}} = 20.2,20.4, 20.6, 20.9$ for the four exposure, schemes assuming available exposures from all position angles. We have also shown that there is an approximate $\Delta m_{\mathrm{F158}} \approx 0.1$ mag decrement corresponding to the $50\%$ efficiency level for each lost exposure. 

Using the derived efficiency curve, we estimate the number density of red galaxies within $1 < z < 2$ that are expected to have reliable identification and spectroscopic redshift measurements in the \textit{Roman} reference HLSS. The direct imaging simulation from \citet{Troxel22}, on which this study is based, covers a region of $20\deg^2$. The number densities of the underlying galaxy populations (e.g., red sequence, blue cloud) used in that simulation are consistent with existing measurements of galaxy luminosity functions~\citep{DC2paper}. We therefore performed a rough estimate by counting the number of red galaxies recorded in the truth catalog of \citet{Troxel22}, then multiplying that by the fraction of galaxies based on the efficiency curve. From this, we predict a total of approximately 60 red, quiescent galaxies per $\deg^2$ within within $1 < z < 2$ that can be identified with accurate redshifts. Extrapolating to the total survey area of the \textit{Roman} HLSS, which spans $2000\ \deg^2$, we predict around 120,000 red, quiescent galaxies in total.
Although such sample consisting solely of red galaxies may not be large enough for analyses such as galaxy clustering and BAO, it can be effectively integrated with other samples to enhance these studies. Moreover, this sample can be utilized for a variety of potential studies. For instance, it can be applied in the RedMaGiC algorithm~\citep{Rykoff14,Rozo16}, which employs a sparse sample of red galaxies to calibrate a red-sequence model. This model improves the accuracy of photometric redshift estimates and facilitates the selection of red galaxy candidates. Additionally, such a sample can contribute to expanding existing high-redshift red galaxy collections, thereby enriching studies of galaxy evolution.

All the data products, including simulated grism images, spectra redshift fitting results and input files required to simulate \textit{Roman} grism with \verb|Grizli| are available upon request.

\begin{acknowledgements}
ZG and CW were supported by Department of Energy, grant DE-SC0010007. 

We thank Yun Wang, Anahita Alavi for sharing necessary \textit{Roman} grism configuration files for simulation. 

We are thankful to Isak G.B. Wood, Debopam Som for discussions on simulation tools.

This work uses \verb|Grizli| package for simulation. We thank Gabriel B. Brammer for their contributions to that package. 

This paper used resources at the Duke Computing Cluster. Plots in this manuscript were produced with MATPLOTLIB~\citep{Hunter07}.
\end{acknowledgements}

\newpage

\bibliography{ref}{}
\bibliographystyle{aasjournal}


\end{CJK*}
\end{document}

%% file: author.tex
\correspondingauthor{Zhiyuan Guo}
\email{zhiyuan.guo@duke.edu}

\author[0000-0001-9557-9171]{Zhiyuan Guo(郭致远)}
\affiliation{Department of Physics, Duke University, Durham NC 27708, USA}

\author[0000-0002-7593-8584]{Bhavin Joshi}
\affiliation{Department of Physics and Astronomy, Johns Hopkins University, Baltimore, MD 21218, USA}

\author[0000-0003-2035-2380]{C. W. Walter}
\affiliation{Department of Physics, Duke University, Durham NC 27708, USA}

\author[0000-0002-5622-5212]{M. A. Troxel}
\affiliation{Department of Physics, Duke University, Durham NC 27708, USA}